Unkel et al. Orphanet Journal of Rare Diseases (2016) 11:16
DOI 10.1186/s13023-016-0402-6Orphanet Journal of
Rare Diseases

RESEARCH  Open Access

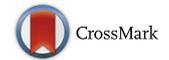

# Systematic reviews in paediatric multiple sclerosis and Creutzfeldt-Jakob disease exemplify shortcomings in methods used to evaluate therapies in rare conditions

Steffen Unkel[1*], Christian Röver[1], Nigel Stallard[2], Norbert Benda[3], Martin Posch[4], Sarah Zohar[5] and Tim Friede[1]**Abstract**

**Background:** Randomized controlled trials (RCTs) are the gold standard design of clinical research to assess interventions. However, RCTs cannot always be applied for practical or ethical reasons. To investigate the current practices in rare diseases, we review evaluations of therapeutic interventions in paediatric multiple sclerosis (MS) and Creutzfeldt-Jakob disease (CJD). In particular, we shed light on the endpoints used, the study designs implemented and the statistical methodologies applied.

**Methods:** We conducted literature searches to identify relevant primary studies. Data on study design, objectives, endpoints, patient characteristics, randomization and masking, type of intervention, control, withdrawals and statistical methodology were extracted from the selected studies. The risk of bias and the quality of the studies were assessed.

**Results:** Twelve (seven) primary studies on paediatric MS (CJD) were included in the qualitative synthesis. No double-blind, randomized placebo-controlled trial for evaluating interventions in paediatric MS has been published yet. Evidence from one open-label RCT is available. The observational studies are before-after studies or controlled studies. Three of the seven selected studies on CJD are RCTs, of which two received the maximum mark on the Oxford Quality Scale. Four trials are controlled observational studies.

**Conclusions:** Evidence from double-blind RCTs on the efficacy of treatments appears to be variable between rare diseases. With regard to paediatric conditions it remains to be seen what impact regulators will have through e.g., paediatric investigation plans. Overall, there is space for improvement by using innovative trial designs and data analysis techniques.

**Keywords:** Clinical trials, Creutzfeldt-Jakob disease, Evidence synthesis, Paediatric multiple sclerosis, Rare diseases, Systematic review# Background

Rare diseases, also referred to as orphan diseases, are illnesses that affect a predefined small proportion of the population. No single cut-off number has been globally agreed upon for which a disease is considered rare and, not surprisingly, various definitions for rare diseases do exist. For example, the European Commission defines the prevalence for a rare disease as affecting no more than 5 per 10 000 persons in the European Union (EU) [1]. It is estimated that at present in the EU, 5000–8000 distinct rare diseases affect 6–8 % of the population, that is, between 27 and 36 million people [2]. Hence, rare diseases create an enormous health and economic burden and research on these deserves considerable attention.

Most rare diseases have a genetic basis and onset is often in childhood and adolescence [2]. Many rare conditions have neurological manifestations [3], of which two prominent examples are paediatric multiple sclerosis (MS) and Creutzfeldt-Jakob disease (CJD). Multiple sclerosis is a disease characterized by recurrent immune-mediated episodes of central nervous system demyelination [4]. Signs and symptoms such as losing the ability to walk independently vary widely, depending on the amount of damage and which areas of the brain are affected. To date there is no cure for MS. However, medications can help speed recovery from attacks of worsening neurologic function, prevent attacks, modify the course of the disease and manage symptoms. Multiple sclerosis has an estimated prevalence of more than 1 individual in 1000 persons and is therefore not rare [5]. Its onset is usually in adult life but in 3–5 % of cases MS manifests before the age of 18 years [6]. Therefore, multiple sclerosis in the subgroup of children and adolescents is rare. The relapse-remitting course is the overwhelming manifestation in early-onset MS with most symptoms of MS similar to those seen in adults [7].

* Correspondence: steffen.unkel@med.uni-goettingen.de
[1]Department of Medical Statistics, University Medical Center Göttingen, Humboldtallee 32, 37073 Göttingen, Germany
Full list of author information is available at the end of the article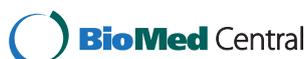

© 2016 Unkel et al. **Open Access** This article is distributed under the terms of the Creative Commons Attribution 4.0 International License (http://creativecommons.org/licenses/by/4.0/), which permits unrestricted use, distribution, and reproduction in any medium, provided you give appropriate credit to the original author(s) and the source, provide a link to the Creative Commons license, and indicate if changes were made. The Creative Commons Public Domain Dedication waiver (http://creativecommons.org/publicdomain/zero/1.0/) applies to the data made available in this article, unless otherwise stated.

Creutzfeldt-Jakob disease is a fatal, untreatable prion encephalopathy. The estimated prevalence of CJD is 1–9 cases per 1,000,000 people [8]. Most frequently, CJD is marked by rapid mental deterioration. The disease evolves into a state of akinetic mutism and ususally leads to death within a few months. Treatment is aimed at alleviating symptoms and making the patient as comfortable as possible [9].

Some major obstacles for expanding knowledge of rare diseases such as paediatric MS and CJD do exist [10, 11]. Firstly, given the small sample size there is lack of clarity on how to design and conduct proper clinical trials to investigate the effects of therapeutic interventions. Whereas in large populations usually two independent randomised controlled trials (RCTs) are required to demonstrate efficacy and safety for marketing authorization, in small populations the conduct of even a single large-scale confirmatory trial might be extremely difficult or not feasible. Besides, if a disease is rare in the subgroup of children or adolescent patients, then one is faced with general ethical and practical challenges of designing and conducting paediatric studies. Secondly, the generation of evidence for rare conditions is often hindered by the lack of knowledge about the clinical course of the disease. Thirdly, the small patient populations can dampen commerical interest in the development of drugs or therapies.

Nevertheless, the past decade has witnessed an increase in research activity on the methodological and statistical issues that are related to the evaluation of interventions in rare diseases and small patient populations. The growth in this area has been so rapid as to spawn conferences, new journals, e.g., the Rare Diseases and Orphan Drugs Journal and the Orphanet Journal of Rare Diseases, and a scientific guideline on clinical trials in small populations published by the European Medicines Agency (EMA) in 2006 [12]. In this regulatory guideline, a range of approaches are discussed that may be helpful at the design stage of small clinical trials such as response-adaptive designs and $n$-of-1 trials. Randomized response-adaptive designs are schemes for patient assignment to treatment, the goal of which is to place more patients on the better treatment based on patient responses already accrued in the trial [13]. A randomized $n$-of-1 study is a clinical trial in which random allocation is used to determine the order in which an experimental and a control intervention are given to a single patient [14].

At the stage of data analysis, sensitivity analyses consisting of various statistical models that make different assumptions about the data should be carried out. The EMA Guideline on clinical trials in small populations also states that the use of Bayesian methods [15] could be advantageous when faced with small datasets because these methods offer a way to combine knowledge from previous data or prior knowledge with data from a study, although introducing prior beliefs is often a concern in drug regulation. As with sensitivity analyses, a variety of reasonable prior distributions should be combined with data from studies to ensure that conclusions are not too much based on a specific prior distribution [12]. The U.S. Food and Drug Administration (FDA) published a draft guidance on rare diseases in August 2015 [16].

Some recently published reviews discuss clinical trial designs and statistical methods that have been proposed or described in the context of rare diseases and small populations [17–19]. More specifically, in [17], clinical trial designs that have been proposed or employed in patients with rare diseases are categorized into two groups: designst that have the potential to reduce the expected trial sample size such as adaptive designs and designs that increase the number or fraction of on-treatment participants such as randomized crossover trials in which subjects receive a sequence of different treatments. Methods for observational data are categorized into four groups: methods to deal with confounding, self-controlled study designs in which patients act as their own controls, case–control designs that involve sampling from an underyling cohort of patients and prospective inception cohorts that are cohorts that are observed starting from a clearly defined point in time such as time of beginning of treatment.

In contrast to [17–19], the purpose of the present paper is to review the methods actually used in the evaluations of therapeutic interventions by considering two prominent rare diseases as examples, namely paediatric MS and CJD. Instead of summarizing the evidence in a narrative fashion, we provide a systematic review that is the result of applying an explicit, rigorous and reproducible logic for identifying and reporting relevant studies.

Although paediatric MS and CJD are clinically very dissimilar, the methodological challenges in evaluating therapeutic interventions in these conditions bear some similarities. Our focus has been on obtaining a qualitative synthesis of the evidence from a methodological perspective. In particular, in view of the paucity of a large number of patients that are available for conducting clinical trials for paediatric MS and CJD, we aim to shed light on the implemented study designs for investigating the endpoints of interest and the statistical methodologies applied to analyse the patient collectives. Our review may offer a way for progressing the study of rare diseases in general.

The remainder of this paper is organized as follows. In Section 2, we describe the literature searches undertaken to identify suitable primary studies for the two illnesses of interest (Section 2.1) and the process of extracting relevant data from those studies (Section 2.2). Section 3 summarizes the main findings; in Section 3.1 and Section 3.2 study selection and study characteristics, respectively, are discussed and in Section 3.3 the methodological quality of the studies is appraised. In Section 4, the review's "take-home" messages are summarized and the strengths and weaknesses of the evidence are discussed. In Section 5, a general interpretation of the results in the context of other evidence is provided and suggestions for future research are made.

## Methods

The preferred reporting items for systematic reviews and meta-analyses (PRISMA) statement [20] guided the writing up of the systematic reviews. The PRISMA checklist of 27 essential items for transparent reporting is given in Additional file 1: Table S1 in the supplementary web material.

### Systematic literature search and article selection

To choose appropriate literature search strategies for the two illnesses under investigation, the PICOS framework [21] was applied. More precisely, we looked for

  i. studies for the subgroup of children or adolescents younger than 18 years of age with a diagnosis of MS and patients with probable or definite CJD, respectively;
 ii. studies with any type of drug intervention;
iii. studies with either no control group or with a control group existing of either placebo concurrent controls, no treatment concurrent controls, dose-comparison concurrent controls (different doses or regimens of same treatment), active treatment concurrent controls (different active treatments), or historical (external and non-concurrent) controls;
 iv. studies with any type of outcome;
  v. all evidence available from meta-analyses, RCTs and controlled/uncontrolled prospective and retrospective observational studies.

Observational studies are deemed to be relevant if at least 20 people were contained in the treatment arm. This imposed restriction eliminates individual case reports and small case series and also may reduce publication bias. Based on these PICOS specifications, publications were identified from the electronic databases PubMed and Cochrane Central Register of Controlled Trials (CENTRAL), by using the following search strategies:

– Paediatric MS: ("Multiple Sclerosis/drug therapy"[MeSH] AND (adolescents[Title/Abstract] OR children[Title/Abstract] OR juvenile[Title/Abstract] OR pediatric[Title/Abstract] OR paediatric[Title/Abstract] OR "early onset"[Title/Abstract])),
– CJD: ("Creutzfeldt-Jakob Syndrome/drug therapy"[MeSH]).

All titles and abtracts were screened independently by two reviewers (SU, CR). Full papers were reviewed if at least one reviewer considered a paper relevant for the review. Finally, the reference lists of all qualifying articles were examined to supplement the search. The primary studies selected were then submitted for data extraction.

### Data extraction, risk of bias assessment tool and quality scales

Using an electronic spreadsheet, the following data items were extracted by one reviewer and verified by another:

- reference,
- aims/objectives of the study,
- study design,
- endpoints of interest,
- description of patient characteristics at the beginning of study,
- method of randomization,
- method of blinding,
- type of intervention and intervention group,
- specification of the control group,
- withdrawals, dropouts, patients lost to follow-up, and
- statistical methodologies used to analyse the patient collectives.

It is important to assess risk of bias in studies as differences in risk of bias can help explain variation in the results of studies included in a systematic review. The Cochrane Collaboration's recommended tool for assessing risk of bias in RCTs is a domain-based evaluation, in which critical assessments are made separately for different domains and then judgements are assigned to relate the risk of bias to each of the domains [22, 23]. This is achieved by assigning a judgement of "low risk", "high risk" or "unclear risk" of bias to the domains covering selection bias, performance bias, detection bias, attrition bias and reporting bias. We have used the Cochrane Collaboration's tool for assessing the risk of bias of the RCTs in patients with paediatric MS and CJD, respectively.

The quality of the selected primary studies has been classified according to the scheme of the American Academy of Neurology (AAN) for assigning levels of evidence for therapeutic questions [24]. The AAN classifies evidence using a four-tiered system (class I through class IV), with class I indicating the strongest evidence and class IV the weakest. The class of evidence is determined both by the rigor of the study and the risk of bias. The criteria used to determine the rigor of the study are similar to those used by other entities that produce guidelines and evidence statements. As with other classification schemes, the presence or absence of randomization and a control group are fundamental elements. A concise summary of the AAN classification scheme with the information relevant for this review is given in Additional file 1: Table S2 in the supplementary material.

In addition, RCTs have been assessed by means of the Oxford Quality Scoring system [25]. The latter scale assesses the quality of published clinical trials based on methods relevant to random assignment, double blinding, and the flow of patients (withdrawals and dropouts). There are seven items. The last two attract a negative score, which implies that the range of possible scores is 0 (very poor) to 5 (rigorous).

The methodological quality of studies included in the review was assessed independently by two researchers (SU, CR). Discrepancies in scoring were resolved through discussion.

## Results

### Study selection

Figures 1 and 2 represent the PRISMA four-phase flow diagrams for the different phases of identification and selection of primary studies on evaluating interventions in paediatric MS and CJD, respectively. Using the search strategies mentioned above, 130 abstracts (124 abstracts) on paediatric MS (CJD) were retrieved from the PubMed and CENTRAL databases. The database searches were conducted on December 15th, 2015.

On the basis of an initial screening of the titles and abstracts, 108 (114) records were excluded on paediatric MS and CJD, respectively. It should be mentioned that for paediatric MS, full-text versions of two records that have been written in Russian and which may have been eligible for inclusion in the review were not available [26, 27]. Scanning the reference lists of the remaining papers revealed a further two (six) papers that may have been eligible for inclusion in the review. After reading the entire manuscripts of twenty-one (thirteen) papers, twelve (seven) primary studies on paediatric MS (CJD) were included in the qualitative synthesis, and nine (six) were excluded with reasons given in Figs. 1 and 2.

### Study characteristics
#### Paediatric MS
Disease-modifying therapies for MS that were successfully tested for adults such as interferon-beta therapy, natalizumab or glatiramer acetate are also tested in patients who are younger than 18 years. An overview about the main characteristics of the twelve selected studies on evaluations of therapeutic interventions in paediatric MS is given in Additional file 1: Table S3. The typical endpoints under investigation are changes in the annualized relapse rate (ARR) and in the expanded disability status scale (EDSS) score. Studies also verify whether the offered treatment is safe and well tolerated in children and adolescent patients. No double-blind, randomized placebo-controlled trial for evaluating interventions in paediatric MS has been published yet. Evidence from one RCT with a total sample size of sixteen patients is available, but the RCT lacks appropriate blinding and the patients in the control group receive no treatment ([28], see also the editorial to the corresponding issue of Journal of Neuropediatrics by A. Minagar).

The eleven observational studies are either uncontrolled treatment versus baseline trials in which patients serve as their own controls or controlled trials including natural history controls, dose-comparison concurrent controls or active treatment concurrent controls. With respect to the analysis of study populations, some studies seem to consider subgroups of patients of the same registry [29–32]. The total sample sizes of the selected studies varies from 16 to 307 (median 52). The RCT [28] has the minimum size of all patient collectives. The number of patients included in the observational studies varies between 20, which is the lower bound imposed by us, and 307.

Comparisons between pre- and post-treatment status or between treatment groups were mostly done by means of statistical hypothesis testing. Except for the RCT [28], no study presented a justification of the sample size on the basis of a power analysis. A power analysis allows determining the sample size required to reject the null hypothesis that there is no treatment effect when the alternative hypothesis is true. Conversely, a power analysis also allows determining the probability of rejecting the null hypothesis that there is no treatment effect when the alternative hypothesis is true, under sample size constraints [33]. The only study that explicitly adjusted for multiple endpoints used the simple Bonferroni correction [26]. Some studies use merely descriptive analyses, e.g., comparing frequencies, to evaluate intervention effects.

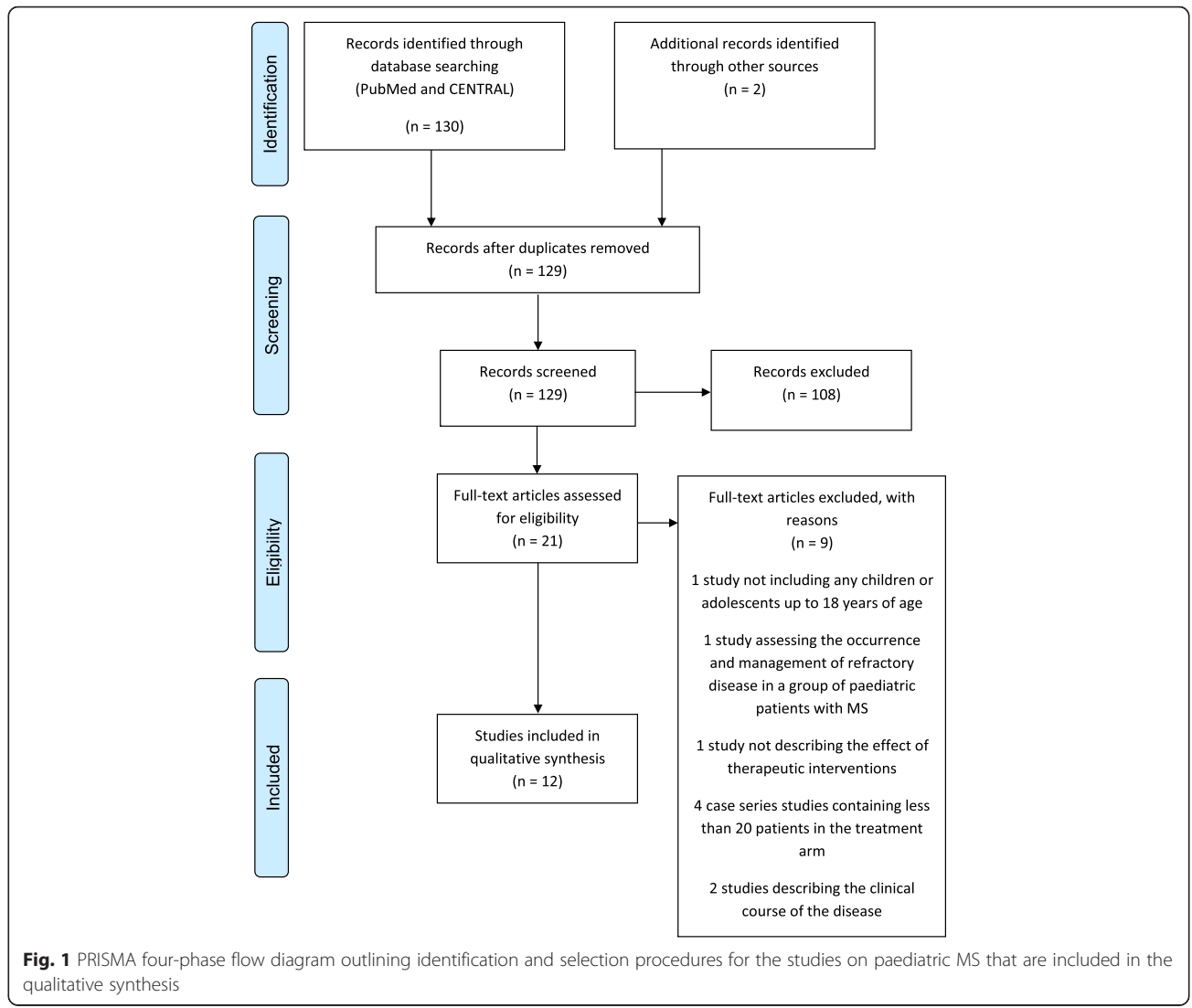

Fig. 1 PRISMA four-phase flow diagram outlining identification and selection procedures for the studies on paediatric MS that are included in the qualitative synthesis

The largest selected study on paediatric MS [34], the study [35] and in part also the study [36] were industry sponsored. With the exception of these three trials, the remaining nine studies appear to be investigator initiated trials. The randomized open-label trial [28] does not provide any information about sponsorship. Apart from the RCT [28], the problem of confounding was addressed only in the comparative observational trial [37], in which regression techniques were applied to eliminate relevant differences between the two treatment groups. Six out of the eleven observational trials are prospective analyses.

### CJD

No therapeutic intervention prevents or reverses the progressive and ultimately fatal course of CJD. The antibiotic doxycycline, the antiviral agent quinacrine and the triaminopyridine compound flupirtine have been tested in the selected studies. Therapies with these drugs aim at alleviating pain and other symptoms and at making people with CJD as comfortable as possible. An overview of the main characteristics of the seven selected studies on evaluations of therapeutic interventions in CJD are given in Additional file 1: Table S4. Not surprsingly, all trials on patients who suffer from CJD investigate time-to-event endpoints, such as survival time from start of treatment to death, time to loss of autonomous feeding, or time to reach the clinical stage of akinetic mutism. In one trial, the primary endpoint was the difference in the cognitive part of the Alzheimer's disease assessment scale (ADAS-cog) [38]. Three of the seven selected studies are double-blind RCTs. One RCT contains a randomized controlled portion and a non-randomized portion [39]; the randomized portion of the trial ended at month 2 and subjects returning for their month-2 visit were offered open-label quinacrine. In one observational study [40], patients were offered a choice between quinacrine, no quinacrine, or randomisation to immediate quinacrine or deferred quinacrine. Four trials are controlled observational studies including natural history controls or historical controls.

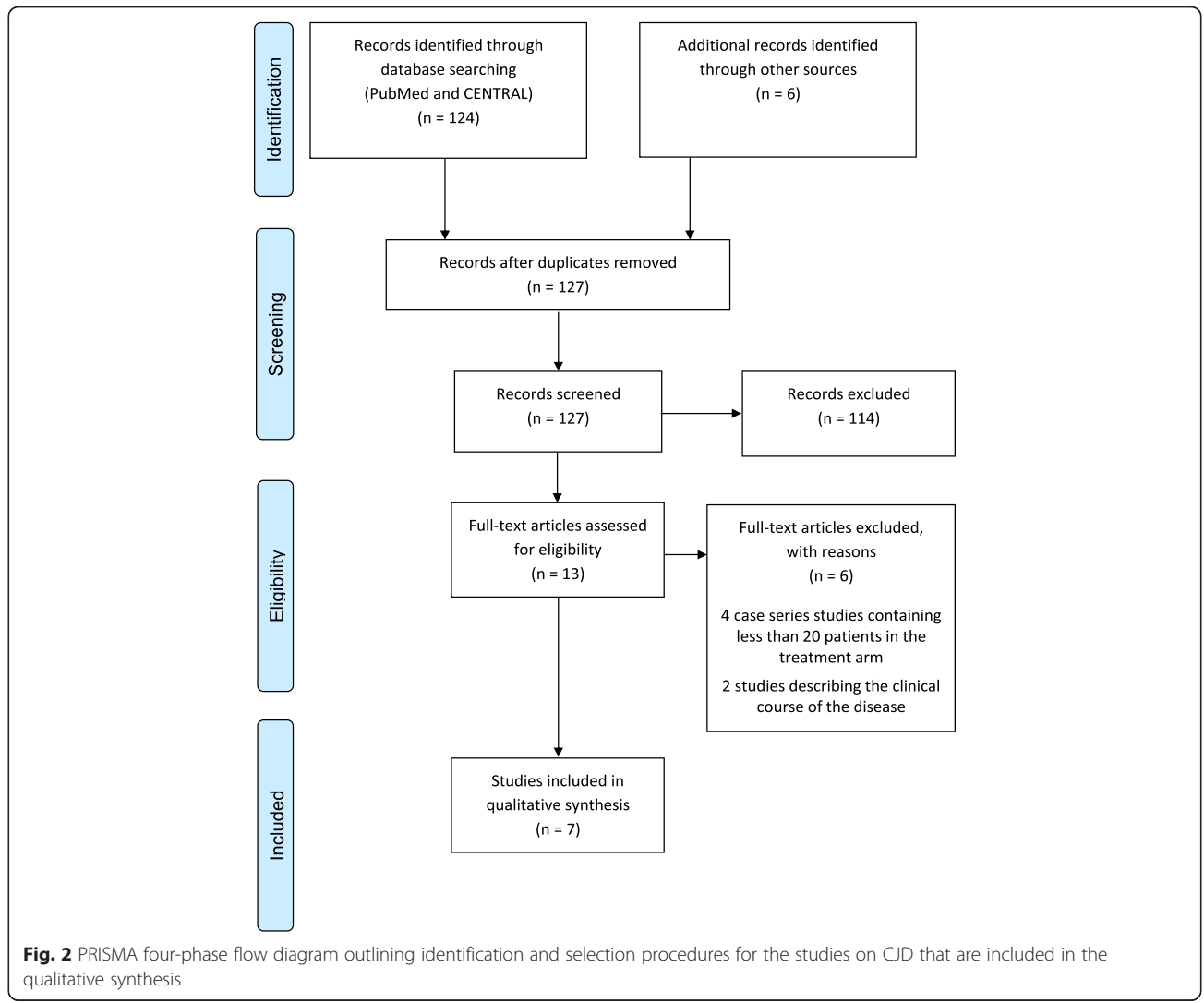

Fig. 2 PRISMA four-phase flow diagram outlining identification and selection procedures for the studies on CJD that are included in the qualitative synthesis

The three RCTs and the patient preference trial presented the sample size obtained from a power analysis [38–41]. Some of the trials only investigate a single primary endpoint. Apparently, in none of the trials that investigate multiple endpoints was any adjustment for multiple testing made. Standard Kaplan-Meier estimates were used to measure the fraction of patients living for a certain amount of time after treatment; they are non-parametric as they do not require specific parametric assumptions to be made about the underlying distribution of survival times [42]. Kaplan-Meier curves for each group of interest were then produced and plotted. The formal comparisons of the survival curves of two groups were based on a formal statistical test. Log-rank tests were employed to test the null hypothesis that there is no difference between the survivor functions of patients in the two treatment groups and to establish the efficacy of an experimental treatment in comparison with a control [42]. Cox proportional hazard models were applied to analyse the effect of covariates on the survival experience of a patient. A Cox regression model provides an estimate of the treatment effect on survival after adjustment for other explanatory variables. In addition, it allows estimating the cumulative risk of an event for an individual, given her or his prognostic variables. The proportional hazards condition requires the ratio of the hazard functions in both treatment groups to remain constant over time [42].

The total sample sizes of the selected studies vary from 28 to 157 (median 99). All studies investigated clinical endpoints such as overall survival or time-to-disease progression. Except for [38], which was partly commerically sponsored, six out of seven studies on CJD are either non-commerically sponsored studies or, as in [43], information about sponsoring was not reported. Two out of the four observational studies are retrospective in nature. The three RCTs address potential confounding by randomization. The open-label patient preference trial [40] adjusts for substantial differences between patients who chose to take quinacrine and those who did not by means of stratification. Table 1 gives a summary of the number of selected studies on paediatric MS and CJD that meet certain methodological characteristics.

### Risk of bias assessment and quality appraisal

Table 2 is a risk of bias summary which is part of the Cochrane Collaboration's tool for assessing risk of bias in RCTs [22]. For each RCT, the judgement (low risk of bias, high risk of bias or unclear risk of bias) is presented for the domains covering selection bias, performance bias, detection bias, attrition bias and reporting bias. The RCTs [39, 41, 44] were judged to have a low risk of bias across all domains indicating a high quality in trial conduct and reporting. Recall that in the trial [39] subjects who were originally randomized were offered open-label treatment at month 2. Therefore, judgements in Table 2 refer to the randomized portion of the trial only. Since in trial [40] patients were offered a choice between treatment, no treatment, or randomisation to immediate treatment or deferred treatment in an open-label, patient-preference trial, this trial is considered an observational study. The single RCT on paediatric MS has methodological flaws and shows a high-risk of performance bias since neither participants nor personnel are blinded to the allocation.

The results of the quality appraisal of the studies on paediatric MS and CJD are presented in Table 3. The evidence on paediatric MS is almost entirely based on observational studies with the quality of studies being classified as class III or class IV. Note that although before-after designs are uncontrolled trials they are categorized in class III on the AAN scheme. In line with [45] some before-after trials were classified as class IV because they lack an independently derived outcome measurement (T. Chitnis, personal communication, May 7th 2015). The single RCT [28] that is available for paediatric MS is ranked class II on the AAN scheme because there is no evidence of concealed allocation; it only received two points on the Oxford Quality Scale because the method of randomisation is not described. Moreover, the study is also not described as double-blind.

For the case of CJD, three of the seven selected studies are RCTs, of which two received the maximum mark on the Oxford Quality Scale [39, 41]. The RCT [38] received four points on this scale because although the study is described as randomised, the method of randomisation is not described. Since the seven selected publications on CJD are either RCTs that fulfill the criteria (a)-(d) stated in Additional file 1: Table S2 or are controlled observational studies, they are classified as class I or class III evidence on the AAN scale.

As indicated in Table 3, the quality of the available evidence for CJD is far better than for paediatric MS. However, the establishment of class I evidence on the effects of therapeutic interventions in the paediatric MS population is hindered by common practical and ethical challenges of studies in children and adolescents. We elaborate further on this point in the Conclusions section.

## Discussion
### Summary of evidence and limitations

Both paediatric MS and CJD have been a focus of great interest in recent years. For these two illnesses there is a clear gap between the methods that have been proposed for studying treatments in small samples and methods that are currently in use. We have identified some shortcomings in the design, analysis and reporting of studies in patients with paediatric MS or CJD. In particular for paediatric MS, the statistical methodology used to analyse the patient collectives is fairly basic, sometimes purely descriptive. One might argue that in studies with few patients there is not much information and so simple analyses are all that are warranted. But regulatory authorities emphasize that, especially for situations when there are very few data, more complex approaches are needed [12]. However, one should be aware that some of the statistical methods that are mentioned in this paper (e.g., hypothesis tests) are based on asymptotic or large sample theory. Within the framework of large sample theory it is typically assumed that the sample size is sufficiently large, and the properties of statistical procedures are evaluated in the limit as the sample size approaches infinity. It may be questioned whether the asymptotic results can be treated as approximately valid for small sample sizes as well.

**Table 1** Number of selected studies on paediatric MS and CJD meeting specific criteria

|  | RCT | Clinical endpoints | Prospective analysis | Sample size calculation | Adjustments for multiple testing | Confounding addressed | Industry sponsored |
|---|---|---|---|---|---|---|---|
| Paediatric MS (n = 12) | 1 | 12 | 6 | 1 | 1 | 2 | 3 |
| CJD (n = 7) | 3 | 7 | 5 | 4 | 0 | 4 | 1 |

**Table 2** Risk of bias assessment of the RCTs on CJD [38, 39, 41] and on paediatric MS [28] (+ low risk of bias, ? unclear risk of bias, - high risk of bias)

| Reference | Random sequence generation (selection bias) | Allocation concealment (selection bias) | Blinding of participants and personnel (performance bias) | Blinding of outcome assessment (detection bias) | Incomplete outcome data (attrition bias) | Selective reporting (reporting bias) |
|---|---|---|---|---|---|---|
| [41] | + | + | + | + | + | + |
| [39] | + | + | + | + | + | + |
| [38] | + | + | + | + | + | + |
| [28] | + | ? | - | ? | + | + |

Almost exclusively, the evidence on therapeutic interventions in paediatric MS found by our review is based on observational studies with the quality of studies being classified as class III or class IV. In most instances, however, observational data should not be relied upon solely to determine whether treatments are efficacious; often the greater biases of observational studies become apparent when their findings are tested with appropriate scientific methods in RCTs. Unfortunately, the single RCT available on patients with paediatric MS suffers from several weaknesses including a very small sample size and lack of blinding. Blinding to subcutaneous interferon-beta-1a treatment could be obtained by allocating matching placebo intramuscular injections to the control group (see for example ClinicalTrials.gov identifier: NCT00441103 for a completed RCT in adult patients with MS).

One promising finding from the RCTs in patients with CJD is that multicentre, occasionally multinational studies, fast enrolment and rigorous study methods with a sufficient number of participants and low risk of bias can be achieved for treatment of a rare, rapidly progressive neurological disorder.

The present review has some limitations. Firstly, our literature searches were focused on retrieving evidence on two specific rare diseases. Because of the large number of rare conditions that do exist, our review is not meant as a comprehensive overview of methods that have been proposed or applied across the spectrum of rare diseases and small populations. Fortunately, there are cases in which statistical methods that have been proposed in the literature for small populations were employed. To give but two examples, a series of $n$-of-1 trials were conducted to evaluate therapies in patients with hereditary angioedema [18], propensity scores were used to match patients with Gaucher disease type I who received different doses of enzyme therapy [17]. Propensity scores allow for a design and an analysis of an observational study that mimic some of the particular characteristics of an RCT [46]. Conditional on the propensity score, the distribution of observed baseline covariates will be similar between treated and untreated subjects.

Secondly, the imposed restriction of considering only studies with at least 20 participants receiving the intervention of interest implies that some evidence for evaluating therapies in paediatric MS and CJD is excluded from this review. On the basis of an initial screening of the titles and abstracts found via the databases PubMed and CENTRAL for paediatric MS (CJD), 18 (13) records out of a total of 130 (124) records, most of which are single case reports, are excluded because they did not meet our imposed restriction.

Finally, we only assessed the risk of bias for RCTs. For controlled studies the domains in the standard tool of the Cochrane Collaboration could also usefully be assessed when the allocation is not randomized [23]. As suggested in [23], for non-randomized studies an additional domain to assess the risk of bias due to confounding should be added. However, since some of selected studies are uncontrolled before-after trials, we decided not to classify the observational studies according to this risk of bias tool. Assessing the extent of the risk of bias of the observational studies that are considered in this review is a topic that is left for further research.

**Table 3** Quality assessment of the studies on paediatric MS (References [28–32, 34–37, 60–62]) and on CJD (References [38–41, 43, 44, 63])

| Studies on paediatric MS | | |
|---|---|---|
| AAN scale | Oxford quality scale for RCTs | References |
| Class II | 2 points | [28] |
| Class III | n/a | [29, 30, 37, 60, 62] |
| Class IV | n/a | [31, 32, 34–36, 61] |
| Studies on CJD | | |
| AAN scale | Oxford quality scale for RCTs | References |
| Class I | 5 points | [39, 41] |
| Class I | 4 points | [38] |
| Class III | n/a | [40, 43, 44, 63] |

### Conclusions

In light of the lack of any class I level evidence on the effects of therapeutic interventions that is available in the paediatric MS population, it would be desirable to have class I level results for this subgroup of patients. This is because an RCT in paediatric patients with MS is essential to reduce the use of medications that are currently off-label and to ensure that children and adolescents are exposed to safe as well as effective treatments. However, trials in the paediatric MS population are confronted with the general practical problem of only a small number of patients available for inclusion. On the one hand, the sample size should be kept as small as possible in order to include no more patients in a trial than necessary with the objective of minimizing unnecessary risks. On the other hand, the sample size should be large enough to ensure statistical power. One way to reduce the trial sample size is to use adaptive designs, of which two types are response-adaptive randomization and sequential trials [47]. Whereas in response-adaptive randomization a greater proportion of patients is assigned to the seemingly more effective treatment while reducing overall trial enrollment, in sequential trials data are analyzed intermittently to guide decisions on termination when safety concerns, futility, efficacy, or a combination of these factors is demonstrated. To achieve the same power for a given treatment effect, studies with an interim analysis have a larger maximum sample size than the fixed sample size design, but the expected sample size, accounting for the gain achieved by early stopping, will typically be smaller for sequential designs than for fixed sample size designs.

Furthermore, ethical challenges of paediatric studies do exist [48]. For example, a placebo-controlled design requires that some participants forego active or possibly effective therapy, which could be a more serious issue in children compared to adults. As a remedy, one might consider alternative designs instead such as add-on studies, active comparator arm studies, dose-ranging studies or deferred-treatment arm studies. A phase III, double-blind, randomised, active-controlled trial to evaluate safety and efficacy of fingolimod versus interferon beta-1a is currently recruiting paediatric patients with MS (ClinicalTrials.gov identifier: NCT01892722). In this RCT, double-dummy masking is required for blinding. Placebo intramuscular injections are obtained by allocating syringes matched in appearance to the active interferon beta-1a intramuscular syringes.

For approval of fingolimod and other pharmacological agents for children and adolescents with MS, paediatric investigation plans are required by the EMA (see for example [49]). These plans are aimed at ensuring the necessary data are obtained through studies in children and adolescents to support the medicine's authorisation for use in this subgroup of patients. In any case, establishing collaborative (e.g., multicentre and multinational) trials should be a priority in rare conditions in general and in rare paediatric populations in particular.

While in general we would call for higher level evidence, there might be settings in which observational studies only could be justified considering the ethical and logistical difficulties in conducting an RCT. In [18], a few examples are presented for which an experimental treatment exists that is markedly better than the standard treatment. In such settings it may be more efficient to run a single-arm trial with all patients receiving the new treatment instead of an RCT. However, sufficient knowledge about the clinical course of the disease is required as this type of marked efficacy would typically be seen only in situations in which an understanding of the disease process has led to a therapy targeted to that process [18].

In situations where randomization is difficult to achieve, methods that incorporate data from other sources in the estimation of the treatment effects may be beneficial. Information external to the study that can be used on the experimental and control arms could stem, for example, from elicitation of experts [50, 51] or related patient populations [52]. Especially in the pediatric MS context a possibility is the extrapolation from adults [53, 54]. If the assumption can be justified that the disease is similar in children and adults and e.g., pre-dominantly older children after puberty are affected that may not be so different to adults, then there may be no need to generate independent level I evidence in children.

Examples where less than the required number of patients are available for randomization include an unbalanced RCT in patients with ankylosing spondylitis [55], an ongoing paediatric study in Alport syndrome [56] and an RCT in patients with early CJD [57]. These three examples may call for the use of methods of generalised evidence synthesis [15, 58], in which studies from different designs are pooled in order to estimate quantities of interest. Various methods for pooling are available [15, 58]. They all share the aim not to eliminate observational studies totally, but rather to provide the information needed to compensate for specific weaknesses of such studies. Generalized evidence synthesis then essentially comes down to making decisions on how and to what extent to discount the observational data. In [55], patients were randomly assigned (in a 4:1 ratio) to either treatment or placebo. For the evaluation of the placebo effect, data from eight previous trials in patients with ankylosing spondylitis were included. The information contained in these trials was transformed into an informative prior distribution by means of the methodology described in [59]. The ongoing trial in paediatric Alport patients aims to combine the treatment effect estimates from the randomized comparison with Alport registry data [56]. A meta-analysis combining evidence on the effects of doxycycline in patients with early CJD from both a randomized study and a non-randomized study is the subject of current research of members of the team of co-authors of this review [57].

## Additional file

**Additional file 1: Supplementary web material.** Supplementary material related to this article is available online. (PDF 93 kb)

**Competing interest**
The authors declare that they have no competing interests.

**Authors' contributions**
S.U. carried out the review, with independent assessement also performed by C.R. Any disagreement was resolved through discussion. T.F. and S.U. were involved in the conception and design of the review. S.U. drafted the manuscript. All authors were involved in the interpretation of the findings, revisions to the original draft and have approved submission of the final manuscript.

**Acknowledgements**
We would like to thank two anonymous reviewers for their helpful suggestions and comments.

**Funding**
This research has received funding from the EU's 7th Framework Programme for research, technological development and demonstration under grant agreement number FP HEALTH 2013 – 602144 with project title (acronym) "Innovative methodology for small populations research" (InSPiRe). The funder had no involvement in the study design; in the collection, analysis and interpretation of the data; in the writing of the report; and in the decision to submit this article for publication.

**Author details**
[1]Department of Medical Statistics, University Medical Center Göttingen, Humboldtallee 32, 37073 Göttingen, Germany. [2]Division of Health Sciences, Warwick Medical School, University of Warwick, Coventry, UK. [3]Biostatistics and Special Pharmacokinetics Unit, Federal Institute for Drugs and Medical Devices, Bonn, Germany. [4]Section of Medical Statistics, Center for Medical Statistics, Informatics and Intelligent Systems, Medical University of Vienna, Vienna, Austria. [5]Institut National de la Santé et de la Recherche Médicale (INSERM), Unité Mixte de Service 1138, Team 22, Centre de Recherche des Cordeliers, Université Paris 5 et Université Paris 6, Paris, France.